\documentclass[aps,rpl,preprint,groupedaddress]{revtex4}
\begin{document}

\title{Teleportation via thermally entangled state of a three-qubit Heisenberg XX ring}
\author{Ye Yeo}

\affiliation{Centre for Mathematical Sciences, Wilberforce Road, Cambridge CB3 0WB, United Kingdom}

\begin{abstract}
We consider quantum teleportation using the thermally entangled state of a three-qubit Heisenberg XX ring as a resource.  Our investigation reveals interesting aspects of quantum entanglement not reflected by the pairwise thermal concurrence of the state.  In particular, two mixtures of different pairs of W states, which result in the same concurrence, could yield very different average teleportation fidelities.
\end{abstract}

\maketitle

Quantum entanglement, as a physical resource, lies at the heart of quantum computation and quantum information \cite{Book}.  An entangled composite system gives rise to nonlocal correlation between its subsystems that does not exist classically.  This nonlocal property enables the uses of local quantum operations and classical communication to teleport an unknown quantum state via a shared pair of entangled particles, with fidelity better than any classical communication protocol \cite{Bennett,Popescu,Horodecki}.  Quantum teleportation can thus serve as an operational test of the presence and strength of entanglement.  It is not only relevant to quantum communication between two distant parties but also to quantum computation, as quantum teleportation is a universal computational primitive \cite{Chuang}.  In Refs.\cite{Karlsson} and \cite{Yeo2}, teleportation of a quantum state using three-particle entangled GHZ state \cite{Greenberger} and W state \cite{Zeilinger} as resources have been demonstrated respectively.  Three-particle entangled states have also been shown to have advantages over the two-particle Bell states in their application to dense coding \cite{Wiesner, Hao} and cloning \cite{Buzek, Ekert}.

In recent years, the presence of entanglement in condensed-matter systems at finite temperatures has been investigated by a number of authors (see, e.g., Refs.\cite{Nielsen, Wang2} and references therein).  The state of a typical condensed-matter system at thermal equilibrium (temperature $T$) is
$\rho = e^{-\beta H}/Z$
where $H$ is the Hamiltonian,
$Z = {\rm tr} e^{-\beta H}$
is the partition function, and
$\beta = 1/kT$
where $k$ is Boltzmann's constant.  The entanglement associated with the thermal state $\rho$ is referred to as thermal entanglement \cite{Arnesen}.  The bulk of these investigations concentrated on the quantification of thermal entanglement and how this quantity changes with temperature $T$ and external magnetic field $B_m$.  More recently, quantum teleportation of an unknown state using the thermally entangled state of a two-qubit Heisenberg XX chain \cite{Wang1} has been demonstrated in Ref.\cite{Yeo1}.  This could have relevance to proposals like the one-way quantum computer \cite{Briegel, Browne}.

In this paper, we consider quantum teleportation in the three-qubit Heisenberg XX ring \cite{Wang2}.  First, we carry out a detailed analysis of the pairwise thermal entanglement in the model, in the presence of an external magnetic field $B_m$.  We find that in contrast to results in Ref.\cite{Wang2}, the antiferromagnetic ring can have nonzero pairwise thermal entanglement when $B_m \not= 0$.  In addition, the maximum amount of pairwise thermal entanglement in the ferromagnetic ring is increased by the presence of $B_m$.  Next, we describe the teleportation scheme $P_1$ \cite{Explanation} and analyze the ``average fidelity criterion'', Eq.(32).  Interestingly, the nonzero thermal entanglement associated with the antiferromagnetic ring cannot yield, for the teleportation scheme $P_1$, average fidelity better than any classical communication protocol.  With the ferromagnetic ring, although quantum teleportation with average fidelity better than any classical communication protocol is possible, the amount of nonzero thermal entanglement does not guarantee this.  In fact, we could have a more entangled thermal state not achieving a better average fidelity than a less entangled one.

The Hamiltonain $H$ for a three-qubit Heisenberg XX ring in an external magnetic field $B_m$ along the $z$ axis is
$$
H = \frac{1}{2} J\left(\sigma^1_A \otimes \sigma^1_B \otimes \sigma^0_C +
                       \sigma^0_A \otimes \sigma^1_B \otimes \sigma^1_C +
                       \sigma^1_A \otimes \sigma^0_B \otimes \sigma^1_C\right. 
$$
$$
\left. + \sigma^2_A \otimes \sigma^2_B \otimes \sigma^0_C +
         \sigma^0_A \otimes \sigma^2_B \otimes \sigma^2_C +
         \sigma^2_A \otimes \sigma^0_B \otimes \sigma^2_C\right)
$$
\begin{equation}
+ \frac{1}{2} B_m \left(\sigma^3_A \otimes \sigma^0_B \otimes \sigma^0_C +
                        \sigma^0_A \otimes \sigma^3_B \otimes \sigma^0_C +
                        \sigma^0_A \otimes \sigma^0_B \otimes \sigma^3_C\right)
\end{equation}
where
$\sigma^0_{\alpha}$
is the identity matrix and
$\sigma^i_{\alpha} (i=1,2,3)$
are the Pauli matrices at site
$\alpha = A, B, C$:
$$
\sigma^0 = \left(\begin{array}{cc}
1 & 0 \\ 0 & 1
\end{array}\right),\ \sigma^1 = \left(\begin{array}{cc}
0 & 1 \\ 1 & 0
\end{array}\right),\ \sigma^2 = \left(\begin{array}{cc}
0 & -i \\ i & 0
\end{array}\right),\ \sigma^3 = \left(\begin{array}{cc}
1 & 0 \\ 0 & -1
\end{array}\right).
$$
$J$ is real coupling constant for the spin interaction.  The ring is said to be antiferromagnetic for $J>0$ and ferromagnetic for $J<0$.  The eigenvalues and eigenvectors of $H$ are given by
$H |000\rangle = \frac{3}{2} B_m |000\rangle$,
$H |W^1\rangle = \frac{1}{2}(B_m + 4J) |W^1\rangle$,
$H |W^{2,3}\rangle = \frac{1}{2}(B_m - 2J) |W^{2,3}\rangle$,
$H |W^4\rangle = -\frac{1}{2}(B_m - 4J) |W^4\rangle$,
$H |W^{5,6}\rangle = -\frac{1}{2}(B_m + 2J) |W^{5,6}\rangle$,
$H |111\rangle = -\frac{3}{2} B_m |111\rangle$, where
$$
|W^1\rangle = \frac{1}{\sqrt{3}}(|001\rangle + |010\rangle + |100\rangle),
$$
$$
|W^2\rangle = \frac{1}{\sqrt{3}}(|001\rangle + q|010\rangle + q^2|100\rangle),
$$
$$
|W^3\rangle = \frac{1}{\sqrt{3}}(|001\rangle + q^2|010\rangle + q|100\rangle),
$$
$$
|W^4\rangle = \frac{1}{\sqrt{3}}(|011\rangle + |101\rangle + |110\rangle),
$$
$$
|W^5\rangle = \frac{1}{\sqrt{3}}(|011\rangle + q|101\rangle + q^2|110\rangle),
$$
$$
|W^6\rangle = \frac{1}{\sqrt{3}}(|011\rangle + q^2|101\rangle + q|110\rangle),
$$
with $q = \exp(i\frac{2}{3}\pi)$.  Here, we use $|0\rangle$ and $|1\rangle$ to denote an orthonormal set of basis states for each two-level system.  For the composite system in equilibrium at temperature $T$, the density operator is
$$
\chi_{ABC} = \frac{1}{Z}
\left[
e^{-\frac{3}{2}\beta B_m} |000\rangle \langle 000| +
e^{-\frac{1}{2}\beta (B_m + 4J)} |W^1\rangle\langle W^1| +
e^{-\frac{1}{2}\beta (B_m - 2J)}  |W^2\rangle\langle W^2| \right.
$$
$$
+ e^{-\frac{1}{2}\beta (B_m - 2J)} |W^3\rangle\langle W^3|
+ e^{\frac{1}{2}\beta (B_m - 4J)} |W^4\rangle\langle W^4|
$$
\begin{equation}
\left. +
e^{\frac{1}{2}\beta (B_m + 2J)}  |W^5\rangle\langle W^5| +
e^{\frac{1}{2}\beta (B_m + 2J)}  |W^6\rangle\langle W^6| +
e^{\frac{3}{2}\beta B_m} |111\rangle \langle 111|
\right]
\end{equation}
where the partition function
$Z = 2\cosh\frac{3}{2}\beta B_m + 2 e^{-2\beta J}\cosh\frac{1}{2}\beta B_m + 4 e^{\beta J} \cosh\frac{1}{2}\beta B_m$,
the Boltzmann's constant $k \equiv 1$ from hereon and
$\beta = 1/T$.
By symmetry under cyclic shifts, the reduced density operators
$\rho_{AB} = {\rm tr}_C\chi_{ABC}$,
$\rho_{BC} = {\rm tr}_A\chi_{ABC}$,
$\rho_{AC} = {\rm tr}_B\chi_{ABC}$
are equal.  In Ref.\cite{Wang2}, Wang {\it et al.} quantify the amount of entanglement associated with $\rho_{AB}$, by considering the concurrence \cite{Wootters, Hill},
$C = \max\{ \lambda_1 - \lambda_2 - \lambda_3 - \lambda_4, 0\}$ where
$\lambda_k (k = 1,2,3,4)$
are the square roots of the eigenvalues in decreasing order of magnitude of the spin-flipped density matrix operator
$R = \rho_{AB} (\sigma^2 \otimes \sigma^2) \rho^{\ast}_{AB} (\sigma^2 \otimes \sigma^2)$,
where the asterisk indicates complex conjugation.  After some straightforward algebra,
$$
\lambda_1 = \frac{2}{3}(2 e^{-2\beta J} + e^{\beta J})\cosh(\frac{1}{2}\beta B_m),\
\lambda_2 = 2e^{\beta J}\cosh(\frac{1}{2}\beta B_m),
$$
\begin{equation}
\lambda_3 = \lambda_4 = \sqrt{\left[\frac{1}{3}(e^{-2\beta J} + 2e^{\beta J})\right]^2 + \frac{2}{3}(e^{-2\beta J} + 2e^{\beta J})\cosh\beta B_m + 1}
\end{equation}
and the thermal concurrence is
\begin{equation}
C(\rho_{AB}) = \max\left\{
\frac{2|e^{-2\beta J} - e^{\beta J}|\cosh\frac{1}{2}\beta B_m - \sqrt{(e^{-2\beta J} + 2e^{\beta J})^2 + 6(e^{-2\beta J} + 2e^{\beta J})\cosh\beta B_m + 9}}{3(\cosh\frac{3}{2}\beta B_m + e^{-2\beta J}\cosh\frac{1}{2}\beta B_m + 2e^{\beta J}\cosh\frac{1}{2}\beta B_m)}, 0
\right\}
\end{equation}
When $B_m = 0$, Eq.(4) reduces to that in Ref.\cite{Wang2}.  In contrast to the two-qubit Heisenberg XX chain \cite{Wang1, Yeo1}, the concurrence is invariant only under the substitution $B_m \rightarrow -B_m$ but not $J \rightarrow -J$.  We thus restrict our considerations to $B_m \geq 0$.  The latter indicates that the entanglement would not be the same for the antiferromagnetic $(J > 0)$ and ferromagnetic $(J < 0)$ cases.

For $J > 0$, Eq.(2) reduces, in the zero temperature limit, i.e., $\beta \longrightarrow \infty$, to the following four possibilities.
\paragraph{Case I: $B_m = 0$}
$$
\chi_{ABC} = \frac{1}{Z}
\left[
|000\rangle \langle 000| + e^{\beta J}(
|W^2\rangle\langle W^2| + |W^3\rangle\langle W^3| +
|W^5\rangle\langle W^5| + |W^6\rangle\langle W^6|) +
|111\rangle \langle 111|
\right]
$$
\begin{equation}
\longrightarrow \frac{1}{4}(
|W^2\rangle\langle W^2| + |W^3\rangle\langle W^3| +
|W^5\rangle\langle W^5| + |W^6\rangle\langle W^6|)
\end{equation}
with $Z = 2 + 4e^{\beta J}$.  From Eq.(4), the above equally-weighted mixture has
\begin{equation}
C(\rho_{AB}) = \max\left\{
-\frac{1 + e^{-2\beta J}}{1 + e^{-2\beta J} + 2e^{\beta J}}, 0
\right\} = 0.
\end{equation}
\paragraph{Case II: $0 < B_m \leq 2J$}
$$
\chi_{ABC} = \frac{1}{Z}
\left[
e^{-\frac{1}{2}\beta (B_m - 2J)}
(|W^2\rangle\langle W^2| + |W^3\rangle\langle W^3|)\right.
$$
$$
\left. + e^{\frac{1}{2}\beta (B_m + 2J)}
(|W^5\rangle\langle W^5| + |W^6\rangle\langle W^6|) +
e^{\frac{3}{2}\beta B_m} |111\rangle \langle 111|
\right]
$$
$$
= \frac{1}{e^{\beta(2B_m - J)} + 2e^{\beta B_m} + 2} (|W^2\rangle\langle W^2| + |W^3\rangle\langle W^3|)
$$
$$
+ \frac{1}{e^{\beta(B_m - J)} + 2 + 2e^{-\beta B_m}} (|W^5\rangle\langle W^5| + |W^6\rangle\langle W^6|)
$$
$$
+ \frac{1}{1 + 2e^{-\beta(B_m - J)} + 2e^{-\beta(2B_m - J)}} |111\rangle \langle 111|
$$
\begin{equation}
\longrightarrow \left\{
\begin{array}{ll}
\frac{1}{2}(|W^5\rangle\langle W^5| + |W^6\rangle\langle W^6|) & {\rm if}\ 0 < B_m < J, \\
\frac{1}{3}(|W^5\rangle\langle W^5| + |W^6\rangle\langle W^6| + |111\rangle\langle 111|) & {\rm if}\ B_m = J, \\
|111\rangle\langle 111| & {\rm if}\ J < B_m \leq 2J
\end{array}
\right.
\end{equation}
with
$Z = e^{\frac{3}{2}\beta B_m} + 2e^{\frac{1}{2}\beta(B_m + 2J)} + 2e^{-\frac{1}{2}\beta(B_m - 2J)}$.  Eq.(4), in the zero temperature limit, reduces to
\begin{equation}
C(\rho_{AB}) = \frac{2}{3(2 + e^{\beta(B_m - J)})} \longrightarrow
\left\{\begin{array}{ll}
\frac{1}{3} & {\rm if}\ 0 < B_m < J, \\
\frac{2}{9} & {\rm if}\ B_m = J, \\
0 & {\rm if}\ J < B_m \leq 2J.
\end{array}\right.
\end{equation}
\paragraph{Case III: $2J < B_m < 4J$}
\begin{equation}
\chi_{ABC} = \frac{1}{Z}
\left[
e^{\frac{1}{2}\beta (B_m + 2J)}
(|W^5\rangle\langle W^5| + |W^6\rangle\langle W^6|) +
e^{\frac{3}{2}\beta B_m} |111\rangle \langle 111|
\right] \longrightarrow |111\rangle\langle 111|
\end{equation}
with
$Z = e^{\frac{3}{2}\beta B_m} + 2e^{\frac{1}{2}\beta(B_m + 2J)}$.
\paragraph{Case IV: $4J \leq B_m$}
$$
\chi_{ABC} = \frac{1}{Z}
\left[
e^{\frac{1}{2}\beta (B_m - 4J)} |W^4\rangle\langle W^4| +
e^{\frac{1}{2}\beta (B_m + 2J)}
(|W^5\rangle\langle W^5| \right.
$$
\begin{equation}
\left. + |W^6\rangle\langle W^6|) +
e^{\frac{3}{2}\beta B_m} |111\rangle \langle 111|
\right] \longrightarrow |111\rangle\langle 111|
\end{equation}
with
$Z = e^{\frac{3}{2}\beta B_m} + e^{\frac{1}{2}\beta(B_m - 4J)} + 2e^{\frac{1}{2}\beta(B_m + 2J)}$.  Clearly, the concurrence in {\it III} and {\it IV} are both zero.  $B_m = J$ therefore marks the point of quantum phase transition from an entangled phase to an unentangled one.

It is obvious from Eq.(6) that in {\it Case I}, the thermal concurrence remains zero even at nonzero temperatures \cite{Wang2}.  However, for $B_m > J$, unequal mixing of entangled and unentangled states in the spectra of the three-qubit Heisenberg XX ring results in nonzero thermal concurrence at $T > 0$.  The thermal entanglement would, in general, decrease in quantity as $\chi_{ABC}$ approaches the maximally mixed state $\frac{1}{8}I_{ABC}$ in the limit of infinite temperature.  It is thus an interesting problem to determine the critical temperatures $T_1$ beyond which the amount of thermal entanglement becomes zero.  From Eq.(4), $T_1$ clearly depends on the external magnetic field $B_m$, in contrast to the two-qubit XX Heisenberg chain \cite{Wang1}.  These $T_1$'s can be obtained by numerically solving
\begin{equation}
2(e^{\beta J} - e^{-2\beta J})\cosh\frac{1}{2}\beta B_m - \sqrt{(e^{-2\beta J} + 2e^{\beta J})^2 + 6(e^{-2\beta J} + 2e^{\beta J})\cosh\beta B_m + 9} = 0
\end{equation}
(see Table I).  We note that $T_1$ increases with increasing $B_m$.  When $B_m$ is large enough, Eq.(11) reduces to
\begin{equation}
x^6 - 6x^5 - 2x^3 - 3x^2 + 1 = 0,
\end{equation}
with $x \equiv e^{\beta J}$, which can be numerically solved to yield $T^*_1 \approx 0.554641J$.  Consequently, as long as $T < T^*_1$, the thermal concurrence would be nonzero, albeit very small.

For $J < 0$, Eq.(2) reduces, in the zero temperature limit, i.e., $\beta \longrightarrow \infty$, to the following four possibilities.
\paragraph{Case V: $B_m = 0$}
$$
\chi_{ABC} = \frac{1}{Z}\left[
|000\rangle \langle 000| +
e^{-2\beta J} (|W^1\rangle\langle W^1| + |W^4\rangle\langle W^4|) +
|111\rangle \langle 111|
\right]
$$
\begin{equation}
\longrightarrow \frac{1}{2}(|W^1\rangle\langle W^1| + |W^4\rangle\langle W^4|)
\end{equation}
with $Z = 2 + 2e^{-2\beta J}$.  From Eq.(4), the above equally-weighted mixture has
\begin{equation}
C(\rho_{AB}) = \max\left\{
\frac{e^{-2\beta J} - (3 + 4e^{\beta J})}{3(1 + e^{-2\beta J} + 2e^{\beta J})}, 0
\right\} \longrightarrow \frac{1}{3}.
\end{equation}
\paragraph{Case VI: $0 < B_m < -2J$}
\begin{equation}
\chi_{ABC} = \frac{1}{Z}
\left[
e^{-\frac{1}{2}\beta (B_m + 4J)} |W^1\rangle\langle W^1| +
e^{\frac{1}{2}\beta (B_m - 4J)} |W^4\rangle\langle W^4| +
e^{\frac{3}{2}\beta B_m} |111\rangle \langle 111|
\right] \longrightarrow |W^4\rangle\langle W^4|
\end{equation}
with $Z = e^{\frac{3}{2}\beta B_m} + e^{\frac{1}{2}\beta(B_m - 4J)} + e^{-\frac{1}{2}\beta(B_m + 4J)}$.  In the zero temperature limit, Eq.(4) reduces to
\begin{equation}
C(\rho_{AB}) = \max\left\{
\frac{2}{3(1 + e^{\beta(B_m + 2J)})}, 0\right\} \longrightarrow \frac{2}{3},
\end{equation}
a signature of $|W^4\rangle$.
\paragraph{Case VII: $-2J \leq B_m \leq -4J$}
$$
\chi_{ABC} = \frac{1}{Z}
\left[
e^{-\frac{1}{2}\beta (B_m + 4J)} |W^1\rangle\langle W^1| +
e^{\frac{1}{2}\beta (B_m - 4J)} |W^4\rangle\langle W^4| \right.
$$
$$
\left. + e^{\frac{1}{2}\beta (B_m + 2J)}
(|W^5\rangle\langle W^5| + |W^6\rangle\langle W^6|) +
e^{\frac{3}{2}\beta B_m} |111\rangle \langle 111|
\right]
$$
\begin{equation}
\longrightarrow \left\{
\begin{array}{ll}
\frac{1}{2}(|W^4\rangle\langle W^4| + |111\rangle\langle 111|) & {\rm if}\ B_m = -2J, \\
|111\rangle\langle 111| & {\rm otherwise}
\end{array}
\right.
\end{equation}
with $Z = e^{\frac{3}{2}\beta B_m} + 2e^{\frac{1}{2}\beta(B_m + 2J)} + e^{\frac{1}{2}\beta(B_m - 4J)} + e^{-\frac{1}{2}\beta(B_m + 4J)}$.  It follows from Eq.(16) and Eq.(17) that
$$
C(\rho_{AB}) = \left\{\begin{array}{ll}
\frac{1}{3} & {\rm if}\ B_m = -2J, \\
0 & {\rm otherwise}.
\end{array}\right.
$$
\paragraph{Case VIII: $-4J < B_m$}
$$
\chi_{ABC} = \frac{1}{Z}
\left[
e^{\frac{1}{2}\beta (B_m - 4J)} |W^4\rangle\langle W^4| +
e^{\frac{1}{2}\beta (B_m + 2J)}
(|W^5\rangle\langle W^5| \right.
$$
\begin{equation}
\left. + |W^6\rangle\langle W^6|) +
e^{\frac{3}{2}\beta B_m} |111\rangle \langle 111|
\right] \longrightarrow |111\rangle\langle 111|
\end{equation}
with $Z = e^{\frac{3}{2}\beta B_m} + 2e^{\frac{1}{2}\beta(B_m + 2J)} + e^{\frac{1}{2}\beta(B_m - 4J)}$.  Consequently, the concurrence is zero in this case.  So, $B_m = -2J$ marks the point of quantum phase transition in the ferromagnetic ring.

The critical temperatures $T_1$ for the ferromagnetic ring again depends on $B_m$ and can be obtained by numerically solving
\begin{equation}
2(e^{-2\beta J} - e^{\beta J})\cosh\frac{1}{2}\beta B_m - \sqrt{(e^{-2\beta J} + 2e^{\beta J})^2 + 6(e^{-2\beta J} + 2e^{\beta J})\cosh\beta B_m + 9} = 0
\end{equation}
(see Table II).  In particular, when $B_m = 0$, Eq.(19) reduces to
\begin{equation}
e^{-2\beta J} - (3 + 4e^{\beta J}) = 0
\end{equation}
which yields $T_1 \approx -1.27136J$ \cite{Wang2}.  We note that $T_1$ similarly increases with increasing $B_m$.  For large enough $B_m$, Eq.(19) reduces to
\begin{equation}
y^6 - 3y^4 - 2y^3 - 6y + 1 = 0,
\end{equation}
with $y \equiv e^{-\beta J}$, which can be numerically solved, giving $T^{**}_1 \approx -1.32639J$.  So, for large $B_m$, the thermal concurrence would be small but nonzero as long as $T < T^{**}_1$.

For both $J > 0$ and $J < 0$, $T_1$ increases with increasing $B_m$ up to $T^*_1$ and $T^{**}_1$ respectively, as long as $B_m$ is not infinitely large, in which case $\chi_{ABC} \longrightarrow |111\rangle$ and the thermal concurrence would become zero.  Physically, one could understand this phenomenon by looking at the thermal density operator, Eq.(2).  Recall that while an equally weighted mixture of $|W^2\rangle$, $|W^3\rangle$, $|W^5\rangle$ and $|W^6\rangle$ has zero concurrence, Eq.(5) and Eq.(6); an equally weighted mixture of $|W^5\rangle$ and $|W^6\rangle$, Eq.(7) and Eq.(8), and $|W^4\rangle$, Eq.(15) and Eq.(16), have nonzero concurrence.  At nonzero temperatures, increasing $B_m$ creates a diminishing proportion of $|W^2\rangle$ and $|W^3\rangle$, but an increasing proportion of $|W^4\rangle$, $|W^5\rangle$, $|W^6\rangle$ (entangled states), and of course $|111\rangle$ (unentangled state).  The small but nonzero proportion of entangled states contributes to the nonzero thermal concurrence.

Now we describe the quantum teleportation protocol $P_1$ using the above three qubit mixed state $\chi_{ABC}$ as a resource.  It involves a sender, Alice (at site $A$), and two receivers, Bob (at $B$) and Cindy (at $C$).  Alice is in possession of two two-level quantum systems, the input system $S$, and another system $A$ entangled with both a third two-level target system $B$ in Bob's possession, and a fourth two-level target system $C$ in Cindy's possession (i.e. a three-particle entangled state).  Here, we label the entangled systems by the site indices.  Initially the composite system $SABC$ is prepared in a state with density operator
$$
\sigma^{total}_{SABC} = \pi_S \otimes \chi_{ABC}
$$
where
\begin{equation}
\pi_S = |\psi\rangle_S\langle\psi|,\
|\psi\rangle_S
= \cos\frac{\theta}{2}|0\rangle_S + e^{i\phi}\sin\frac{\theta}{2}|1\rangle_S,
\end{equation}
$0 \leq \theta \leq \pi,\ 0 \leq \phi \leq 2\pi$ are the polar and azimuthal angles respectively, and $\chi_{ABC}$, is as given in Eq.(2).  To teleport the input state $\pi_S$ to Bob's target system $B$ and Cindy's target system $C$, Alice performs a joint Bell basis measurement on systems $S$ and $A$, described by operators $\Pi^j_{SA} \otimes I_{BC}$, $I_{BC}$ is the identity operator on the composite subsystem $BC$, $j$ labels the outcome of the measurement,
\begin{equation}
\Pi^1_{SA} = |\Phi^+\rangle_{SA}\langle\Phi^+|,\
\Pi^2_{SA} = |\Phi^-\rangle_{SA}\langle\Phi^-|,\
\Pi^3_{SA} = |\Psi^+\rangle_{SA}\langle\Psi^+|,\
\Pi^4_{SA} = |\Psi^-\rangle_{SA}\langle\Psi^-|,
\end{equation}
where
$$
|\Phi^{\pm}\rangle_{SA} = \frac{1}{\sqrt{2}}(|00\rangle_{SA} \pm |11\rangle_{SA}),
$$
$$
|\Psi^{\pm}\rangle_{SA} = \frac{1}{\sqrt{2}}(|01\rangle_{SA} \pm |10\rangle_{SA})
$$
are the Bell states.  If Alice's measurement has outcome $j$, she broadcasts her measurement result (two-bit) to Bob and Cindy via a classical channel.  The joint state of Bob's target system $B$ and Cindy's target system $C$ conditioned on Alice's measurement result $j$ is given by
\begin{equation}
\rho^j_{BC} = \frac{1}{p_j}{\rm tr}_{SA}[(\Pi^j_{SA} \otimes I_{BC})(\pi_S \otimes \chi_{ABC})],
\end{equation}
where
\begin{equation}
p_j = {\rm tr}_{SABC}[(\Pi^j_{SA} \otimes I_{BC})(\pi_S \otimes \chi_{ABC})].
\end{equation}
Substituting Eq.(2), Eq.(22), and Eq.(23) into Eq.(25) yields
$$
p_1 = p_2 = \frac{1}{12Z}e^{-2\beta(B_m + J)}(f + g\cos\theta),
$$
\begin{equation}
p_3 = p_4 = \frac{1}{12Z}e^{-2\beta(B_m + J)}(f - g\cos\theta),
\end{equation}
where
$$
f = 3e^{\frac{3}{2}\beta B_m}(1 + e^{\beta B_m})(1 + 2e^{3\beta J}) + 3e^{\frac{1}{2}\beta B_m}(1 + e^{3\beta B_m})e^{2\beta J},
$$
$$
g = e^{\frac{3}{2}\beta B_m}(1 - e^{\beta B_m})(1 + 2e^{3\beta J}) + 3e^{\frac{1}{2}\beta B_m}(1 - e^{3\beta B_m})e^{2\beta J}.
$$
For Bob and Cindy to successfully complete the teleportation protocol, they perform $j$-dependent unitary operations $U^j_B = U^j_C = U^j$ on systems $B$ and $C$ respectively $(\rho^j_B = {\rm tr}_C\rho^j_{BC} = {\rm tr}_B\rho^j_{BC} = \rho^j_C = \rho^j)$ such that
\begin{equation}
\tau^j_B = \tau^j_C = \tau^j = U^j\rho^jU^{j\dagger},
\end{equation}
where $U^j$ could either be the identity matrix or one of the Pauli matrices (see Table III).  The success of the teleportation scheme can be measured by the fidelity \cite{Jozsa} between the input state $\pi_{in}$ and the output state $\tau^j_{out}$, averaged over all possible Alice's measurement outcomes $j$ and over an isotropic distribution of input states $\pi_{in}$:
\begin{equation}
\langle F\rangle = 
\frac{1}{4\pi}\int^{\pi}_0\int^{2\pi}_0\sin\theta d\theta d\phi\
\sum^4_{j = 1}p_jF^j
\end{equation}
where
\begin{equation}
F^j \equiv {\rm tr}(\tau^j_{out}\pi_{in}).
\end{equation}
It follows from Eq.(22) and results from Eq.(24) that
$$
F^1 = F^2 = \frac{h_1 + h_2\cos2\theta}{4(f + g\cos\theta)},
$$
\begin{equation}
F^3 = F^4 = \frac{h_1 + h_2\cos2\theta}{4(f - g\cos\theta)},
\end{equation}
where
$$
h_1 = 3e^{\frac{3}{2}\beta B_m}(1 + e^{\beta B_m})(3 + 4e^{3\beta J}) + 3e^{\frac{1}{2}\beta B_m}(1 + e^{3\beta B_m})e^{2\beta J},
$$
$$
h_2 = -e^{\frac{3}{2}\beta B_m}(1 + e^{\beta B_m})(1 - 4e^{3\beta J}) - 3e^{\frac{1}{2}\beta B_m}(1 + e^{3\beta B_m})e^{2\beta J}.
$$
Substituting Eq.(26) and Eq.(30) into Eq.(28) gives
\begin{equation}
\langle F\rangle = \frac{1}{3} + \frac{2}{9}\frac{(2 + e^{3\beta J})\cosh\frac{1}{2}\beta B_m}{(1 + 2e^{3\beta J})\cosh\frac{1}{2}\beta B_m + e^{2\beta J}\cosh\frac{3}{2}\beta B_m}
\end{equation}
In order to transmit $\pi_{in}$ with fidelity better than any classical communication protocol, we require $\langle F\rangle$ to be strictly greater than $\frac{2}{3}$.  In other words, we require
\begin{equation}
\frac{1}{3}(e^{-2\beta J} - 4e^{\beta J}) > \frac{\cosh\frac{3}{2}\beta B_m}{\cosh\frac{1}{2}\beta B_m}
\end{equation}
and hence $J < 0$.  That is, the nonzero thermal entanglement for $J > 0$ is ``not suitable'' as a resource for teleportation via $P_1$.

In the zero temperature limit, Eq.(31) reduces to
\begin{equation}
\langle F\rangle = \frac{1}{3} + \frac{4}{9(1 + e^{\beta(B_m + 2J)})} \longrightarrow\left\{\begin{array}{ll}
\frac{7}{9} & {\rm if}\ 0 \leq B_m < -2J, \\
\frac{5}{9} < \frac{2}{3} & {\rm if}\ B_m = -2J, \\
\frac{1}{3} < \frac{2}{3} & {\rm if}\ -2J < B_m.
\end{array}\right.
\end{equation}
So, in spite of the fact that in {\it Case V}, the concurrence is only $\frac{1}{3}$, the equally weighted mixture in Eq.(13) is able to yield $\langle F\rangle = \frac{7}{9}$.  Comparing this with the equally weighted mixture  of $|W^5\rangle$ and $|W^6\rangle$ in Eq.(7), which has concurrence also equal to $\frac{1}{3}$, but cannot yield $\langle F\rangle > \frac{2}{3}$, certainly illustrates a fundamental difference between the entangled mixed states not reflected by the concurrence.  For $0 < B_m < -2J$, $\langle F\rangle = \frac{7}{9}$ is a clear signature of $|W^4\rangle$ (see Ref.\cite{Yeo2}).  At $B_m = -2J$, the mixing of $|W^4\rangle$ with an equal proportion of $|111\rangle$ deteriorates the ``quality'' of the entanglement so much that $\langle F\rangle$ is now less than $\frac{2}{3}$.  We note that $B_m = -2J$ marks the point of ``transition'' from $\langle F\rangle > \frac{2}{3}$ to $\langle F\rangle \leq \frac{2}{3}$.  This coincides with the point of quantum phase transition in the ferromagnetic ring.

For nonzero temperatures it is again an interesting problem to determine the critical temperatures $T_2$ beyond which $\langle F\rangle \leq \frac{2}{3}$.  From Eq.(32), $T_2$ is clearly dependent on the magnetic field $B_m$, as in Ref.\cite{Yeo1}.  They can be obtained by numerically solving
\begin{equation}
(e^{-2\beta J} - 4e^{\beta J})\cosh\frac{1}{2}\beta B_m - 3\cosh\frac{3}{2}\beta B_m = 0.
\end{equation}
(see Table II).  Interestingly, when $B_m = 0$, Eq.(34) reduces to Eq.(20) which thus yields $T_2 = T_1 \approx -1.27136J$.  This means that all nonzero thermal entanglement, in this case, is ``suitable'' as a resource for teleportation via $P_1$.  The mixing of states here clearly does not have a devastating effect on the quality of the thermal entanglement.  Supppose $T$ is small enough, and let $B_m = -\eta J$, $0 < \eta$, then Eq.(34) yields
\begin{equation}
T_2 = -\frac{1}{\ln 3}(2 - \eta)J \geq 0.
\end{equation}
So, $\eta$ can at most equal 2, and as $\eta \longrightarrow 2$, $T_2 \longrightarrow 0$ consistent with our assumption.  As shown in Table II, $T_2$ decreases with increasing $B_m$ and each $T_2$ is strictly less than the corresponding $T_1$, which increases asymptotically to $T^{**}_1$.  This means that with increasing $B_m$ we have an increasing range of nonzero thermal entanglement which is however not able to yield $\langle F\rangle > \frac{2}{3}$.  Physically, one could attribute the cause of the poor quality of thermal entanglement to the fact that there is now a comparable or greater proportion of unentangled $|111\rangle$ than the ``teleportation grade'' $|W^1\rangle$ and $|W^4\rangle$.

In conclusion, our ``average teleportation fidelity criterion'', Eq.(32) reveals several interesting aspects of quantum entanglement not reflected by concurrence.  On the one hand, whereas comparable mixing of entangled states certainly decreases the resulting pairwise concurrence (see Eq.(5) and Eq.(6), Eq.(7) and Eq.(8), Eq.(13) and Eq.(14)), it could result either in low quality states which yield $\langle F\rangle \leq \frac{2}{3}$, or in high quality states giving $\langle F\rangle > \frac{2}{3}$.  On the other hand, comparable mixing of entangled with unentangled states not only certainly decreases the resulting pairwise concurrence (see Eq.(7) and Eq.(8), Eq.(17)) but definitely degrades the teleportation quality of the entangled mixed state.  Furthermore, the teleportation quality of the entangled mixed state is more sensitive to the degree of mixing than its concurrence.  As a result, we could have a more entangled thermal state not giving a better average fidelity than a less entangled one.  Since entanglement is such an important resource in quantum information, it is very important to have a more fundamental understanding of these aspects of quantum entanglememt.\\

The author thanks Yuri Suhov and Andrew Skeen for useful discussions.  This publication is an output from project activity funded by The Cambridge MIT Institute Limited (``CMI'').  CMI is funded in part by the United Kingdom Government.  The activity was carried out for CMI by the University of Cambridge and Massachusetts Institute of Technology.  CMI can accept no responsibility for any information provided or views expressed.

\newpage

\begin{table}
\begin{ruledtabular}
\begin{tabular}{cl}
$\eta$ & $T_1$ \\
0.1 & 0.234194J \\
0.3 & 0.332167J \\
0.6 & 0.414045J \\
1.0 & 0.476533J \\
1.3 & 0.504831J \\
2.0 & 0.538225J \\
7.0 & 0.554639J \\
9.0 & 0.554641J \\
10.0 & 0.554641J \\
100.0 & 0.554641J
\end{tabular}
\end{ruledtabular}
\caption{\label{I}}
The critical temperature $T_1$ is a function of both $J$ and $B_m = \eta J,\ \eta > 0$.
\end{table}

\begin{table}
\begin{ruledtabular}
\begin{tabular}{cll}
$\eta$ & $T_1$ & $T_2$\\
0.0 & -1.27136J & -1.27136J \\
0.6 & -1.27457J & -1.17224J \\
0.8 & -1.27686J & -1.08726J \\
1.0 & -1.27959J & -0.965516J \\
1.2 & -1.28263J & -0.795176J \\
1.4 & -1.28585J & -0.578739J \\
1.6 & -1.28916J & -0.368014J \\
1.8 & -1.29246J & -0.182056J \\
1.9 & -1.29408J & -0.0910239J \\
2.0 & -1.29567J & 0 \\
10.0 & -1.32628J & - \\
15.0 & -1.32639J & - \\
16.0 & -1.32639J & - \\
100.0 & -1.32639J & -
\end{tabular}
\end{ruledtabular}
\caption{\label{II}}
The critical temperatures $T_1$ and $T_2$ are functions of both $J$ and $B_m = -\eta J,\ \eta > 0$.
\end{table}

\begin{table}
\begin{ruledtabular}
\begin{tabular}{ccc}
Alice's measurement result $j$ & Bob's unitary operation $U^j$ & Cindy's unitary operation $U^j$ \\
1 & $\sigma_x$ & $\sigma_x$\\
2 & $\sigma_y$ & $\sigma_y$\\
3 & $I$ & $I$\\
4 & $\sigma_z$ & $\sigma_z$
\end{tabular}
\end{ruledtabular}
\caption{\label{III}}
Bob's and Cindy's unitary operations conditioned only on  Alice's measurement results.
\end{table}


\begin{thebibliography}{99}
\bibitem{Explanation} In Ref.[6], Karlsson and Bourennane considered two teleportation schemes $P_0$ and $P_1$.  $P_0$ involves a sender A, an accomplice C, and a receiver B.  A performs a joint Bell basis measurement, C a single-qubit measurement, and B a unitary operation conditioned on A and C's measurement outcomes.  It is shown in Ref.[7] that for three-particle entangled W state, $P_0$ and $P_1$ yield the same average teleportation fidelity.  So, we consider the simpler $P_1$ here.
\end{thebibliography}
\end{document}